%% file: template.tex
\documentclass{article}

\usepackage{arxiv}

\usepackage{graphicx}%
\usepackage{multirow}%
\usepackage{amsmath,amssymb,amsfonts}%
\usepackage{amsthm}%
\usepackage{mathrsfs}%
\usepackage[title]{appendix}%
\usepackage{xcolor}%
\usepackage{textcomp}%
\usepackage{manyfoot}%
\usepackage{booktabs}%
\usepackage{algorithm}%
\usepackage{algorithmicx}%
\usepackage{algpseudocode}%
\usepackage{listings}%
\usepackage{bm}
\usepackage{tabularx}

\input{math_commands}

\newtheorem{definition}{Definition}[section]

\title{Material Property Prediction using Graphs based on Generically Complete Isometry Invariants}

\author{
 Jonathan Balasingham \\
  Department of Computer Science\\
University of Liverpool\\
Liverpool L69 3BX, UK\\
  \texttt{jbalasin@liverpool.ac.uk} \\
   \And
 Viktor Zamaraev \\
  Department of Computer Science\\
University of Liverpool\\
Liverpool L69 3BX, UK\\
  \texttt{Viktor.Zamaraev@liverpool.ac.uk} \\
  \And
 Vitaliy Kurlin \\
  Department of Computer Science\\
University of Liverpool\\
Liverpool L69 3BX, UK\\
  \texttt{Vitaliy.Kurlin@liverpool.ac.uk} 
}

\newcommand{\CG}{\mathrm{CG}}
\newcommand{\PDD}{\mathrm{PDD}}
\newcommand{\DDG}{\mathrm{DDG}}
\newcommand{\angstrom}{\textup{\AA}}
\newcommand{\Z}{\mathbb{Z}}

\begin{document}
\maketitle
\begin{abstract}
The structure-property hypothesis says that the properties of all materials are determined by an underlying crystal structure.
The main obstacle was the ambiguity of conventional crystal representations based on incomplete or discontinuous descriptors that allow false negatives or false positives.
This ambiguity was resolved by the ultra-fast Pointwise Distance Distribution (PDD), which distinguished all periodic structures in the world's largest collection of real materials (Cambridge Structural Database).       
The state-of-the-art results in property predictions were previously achieved by graph neural networks based on various graph representations of periodic crystals, including the Crystal Graph with vertices at all atoms in a crystal unit cell. 
This work adapts the Pointwise Distance Distribution for a simpler graph whose vertex set is not larger than the asymmetric unit of a crystal structure. The new Distribution Graph reduces mean-absolute-error by 0.6\%-12\% while having 44\%-88\% of the number of vertices when compared to the crystal graph when applied on the Materials Project and Jarvis-DFT datasets using CGCNN and ALIGNN. Methods for hyper-parameters selection for the graph are backed by the theoretical results of the Pointwise Distance Distribution and are then experimentally justified.
\end{abstract}


\section{Introduction}
\label{sec:intro}





Traditionally, material properties are predicted using computationally expensive simulation methods like Density Functional Theory (DFT). DFT (and other similar methods) are prohibitively slow and require the tuning of many parameters for specific chemical compositions \cite{dft1}. 
Furthermore, DFT calculations introduce their own approximation error due to the relaxation of constraints required to make the computations feasible \cite{cohen2012challenges}. 
These reasons have led to the application of machine learning algorithms instead, to decrease computation time and allow for more general application on materials. 
\smallskip

Early property prediction methods used kernel regression \cite{LR1}, feed-forward neural networks \cite{deepANN}, and ensemble methods \cite{ensemble}.
Since molecules and crystals have underlying structures, the more recent
Graph Neural Networks (GNN) \cite{seminalGNN} encode the information for atoms and bonds in the vertices and edges of a graph.
\smallskip
 
Crystal structures are much more complex than finite molecules due to their periodic nature, leading to ambiguous representations \cite{widdowson2022resolving}. 
One of the most prominent contributions to crystal property prediction was the Crystal Graph Convolutional Neural Network (CGCNN) \cite{CGCNN} based on 
the \textit{Crystal Graph} incorporating inter-atomic interactions up to a manually chosen cut-off radius. 
The Crystal Graph has been used in many material property predictions \cite{rep_learning, attentionGNN,CGAT, MT_CGCNN, deeperGNN, crysXPP, liu2023symmetryinformed} including the recent state-of-the-art models ALIGNN \cite{ALIGNN} and Matformer\cite{matformer} for crystals in the Materials Project \cite{MP1,matbench}. 
Models that extend CGCNN incorporate molecular representations \cite{graphormer} include three-body interactions between atoms \cite{iCGCNN}, define edges as bonds with a cutoff radius \cite{MEGNET}, or use specialized attention masks to add geometric data \cite{geoCGCNN}. 

All the advances described above modified the algorithm, while the ambiguity of input crystal data remained unresolved because any crystal structure has infinitely many different unit cells.
Doubling a unit cell preserves the underlying crystal structure but doubles the size of graph-based representations.
Even a minimal (primitive) unit cell can be chosen in infinitely many ways related by transformations from the group SL$(\mathbb{Z}^n)$ that change a linear basis, but keep the same periodic lattice. Efforts in crystallography tried to reduce the ambiguity of crystal representations by choosing a reduced cell, which has led to a deeper discontinuity problem that was not obvious many years ago when a few experimental crystals were compared by manual methods.
The best-known Niggli's cell \cite{lawton1965reduced} has been known to be experimentally discontinuous under atomic perturbations since 1965.
\smallskip

\begin{figure}[h!]
\centering
\includegraphics[width=\linewidth]{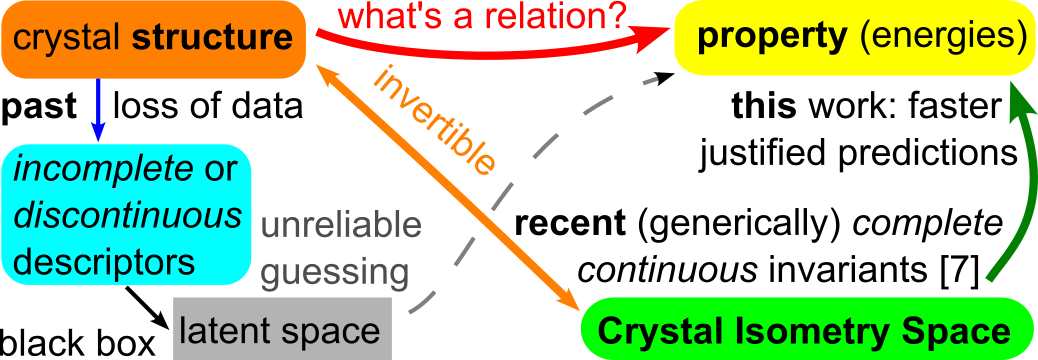}
\caption{Past work used incomplete or discontinuous representations of crystals.
This work predicts energies faster by adapting generically complete and continuous invariants \cite{widdowson2022resolving}.}
\label{fig:structure-property-relations}
\end{figure}

Experimental coordinates of unit cells and atoms are inevitably affected by atomic vibrations and measurement noise.
Crystal Structure Prediction tools output millions of simulated crystals \cite{Pulido2017-ff} as iterative approximations to local minima of complicated energy.
Many of these approximations accumulate around the same local minimum but their vastly different representations were not recognized as near-duplicates. 
Fig.~\ref{fig:structure-property-relations} summarizes the property prediction process using incomplete or discontinuous descriptors that ambiguously map crystals to a latent space, now replaced by the \emph{Crystal Isometry Space}  \cite{widdowson2022resolving} of isometry classes of all periodic crystals. 

The present work introduces a simpler graph representation of a periodic crystal called the \textit{Distance Distribution Graph} (DDG) that is independent of a primitive unit cell and invariant under rigid motion. 
The new graph is significantly smaller than the Crystal Graph in terms of vertices and edges, making it more computationally efficient both in terms of memory and speed. 
Tables~\ref{tab:bycoltolmp} and~\ref{tab:bycoltolt2} will show that the DDG outperforms the Crystal Graph on the Materials Project crystals \cite{MP1} and the T2 dataset of simulated crystals \cite{Pulido2017-ff}. 
We will experimentally show how the tuning of the hyper-parameters affects outcomes, and provide guidance on their selection based on practical and theoretical results. 

\section{Methodology: invariant graphs}
\label{sec:methodology}

We justify a new invariant-based approach to periodic crystals in Subsection~\ref{sub:periodic} and then adapt a generically complete matrix invariant for a new graph invariant in Subsection~\ref{sub:PDD} for comparison with past graph representations by using Graph Neural Networks in Subsection~\ref{sub:GNN}.

\subsection{Isometry invariants of periodic sets of points}
\label{sub:periodic}

Any periodic crystal can be formally defined as a periodic set of points at all atomic centers \cite{widdowson2022average}.
\smallskip

\begin{definition}[Periodic Point Set $S=M+\Lambda\subset\mathbb{R}^n$]
\label{dfn:periodic_set}
For any basis vectors $\mathbf{v}_1 \ldots \mathbf{v}_n$ of $\mathbb{R}^n$,
the \emph{lattice} $\Lambda\subset\mathbb{R}^n$ is formed by all linear combinations $\sum_{i=1}^n c_i v_i$ with  integer coefficients $c_i\in \mathbb{Z}$. 
Considering all $c_i$ in the half-open interval $[0, 1)$, we obtain the \emph{unit cell} $U$ of this basis. A \emph{motif} $M\subset U$ is a finite set of points. A \emph{periodic point set} is $S=M+\Lambda=\{ \bm{p}+\bm{\lambda} \mid \bm{p}\in M,\bm{\lambda} \in \Lambda \}$.
\end{definition}
\smallskip

Any lattice can have infinitely many different bases (hence, unit cells), see Fig.~1 of \cite{widdowson2022resolving}.
Even if we fix a basis, different motifs can generate periodic point sets that are related by rigid motion (or isometry) defined below.
\smallskip

\begin{figure}
\centering
\includegraphics[width=0.9\linewidth]{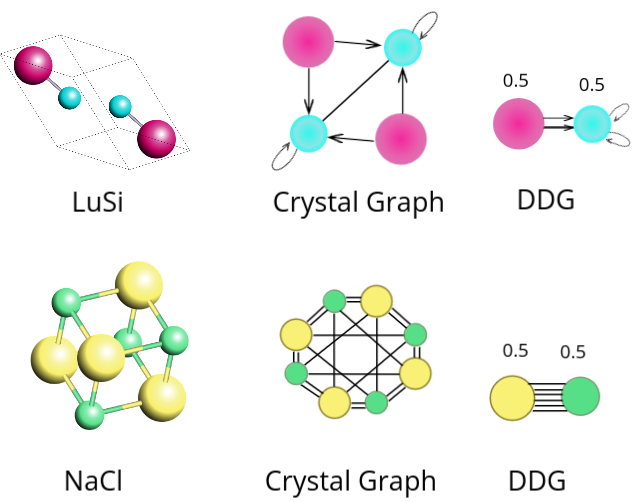}
\caption{(\textit{top}) The transformation from the crystal structure of LuSi to the Crystal Graph and DDG with $k=2$ nearest neighbors. (\textit{bottom}) The transformation from the crystal structure of NaCl to the Crystal Graph and DDG using $k=4$ and the conventional unit cell (edges without arrowhead indicate bidirectionality).}
\label{fig:LuSi}
\end{figure}

Recall that an \emph{orientation} of $\mathbb{R}^n$ can be defined as the sign of the determinant of the $n\times n$ matrix whose columns $v_1,\dots,v_n$ form a linear basis of $\mathbb{R}^n$.

\smallskip
\begin{definition}[Isometry Invariants]
\label{dfn:isometry}
An \emph{isometry} is any map $f:\mathbb{R}^n\to\mathbb{R}^n$ that preserves Euclidean distances. Any isometry $f$ decomposes into translations, rotations, and reflections.
If we exclude reflections, $f$ preserves an orientation of $\mathbb{R}^n$ and can be included in a \emph{rigid motion}, which is a continuous family of isometries $f_t:\mathbb{R}^n\to\mathbb{R}^n$, $t\in[0,1]$, connecting $f_1=f$ with the identity map $f_0$. An isometry \emph{invariant} of periodic set $S$ is a function $I(S)$ such that if $S\simeq Q$ are isometric, then $I(S)=I(Q)$, so $I$ has \emph{no false negatives} such that there are pairs $S\simeq Q$ with $I(S)\neq I(Q)$. 
An invariant $I$ is \emph{complete} if the converse holds: if $I(S)=I(Q)$, then $S\simeq Q$ are isometric, so $I$ has \emph{no false positives} that are pairs $S\not\simeq Q$ with $I(S)=I(Q)$. 
\end{definition}

\smallskip
To be useful, an invariant $I$ should have easily comparable values.
Otherwise, one can define the complete invariant $I(S)$ equal to the isometry class of $S$, which consists of all (infinitely many) isometric images of $S$. Any constant such as $I(S)=0$ is also invariant, so helpful invariants are non-constant.
If $I(S)\neq I(Q)$, the invariance property implies that $S\not\simeq Q$ are not isometric. 
Any non-invariant cannot guarantee this conclusion, hence only invariants can reliably distinguish crystals.

Since points of a periodic set $S$ in Def.~\ref{dfn:periodic_set} are not labeled, similarly to identical atoms in real crystals such as carbons in graphite or diamond, an invariant in the sense of Def.~\ref{dfn:isometry} should be also preserved under all permutations of points.
The physical density of a crystal (or point density of a periodic point set) is an isometry invariant but is incomplete because many different crystals have the same (or nearly equal) densities.
\smallskip

A complete invariant is like the DNA code, which uniquely identifies any human in practice, except for identical twins, which can be considered a singular case, so we could call the DNA a generically complete invariant that distinguishes almost all living organisms.
\smallskip

The past complete invariants of periodic crystals were usually based on a reduced cell of an underlying lattice, which is discontinuous even in dimension 1.
Indeed, the sequence $\{0,1+\varepsilon_1,\dots,m+\varepsilon_m\}+(m+1)\Z$ is nearly identical to the set $\Z$ of integers for all $\varepsilon_i$ close to $0$, but the periods $1$ and $1+m$ (unit cells) are arbitrarily different.
This discontinuity was resolved by the complete and continuous invariant \emph{isoset} \cite{anosova2021isometry} whose disadvantage was an approximate algorithm for a metric computation \cite{anosova2022algorithms}, which has a guaranteed multiplicative factor of about 4 in $\mathbb{R}^3$.
The next subsection discusses the much faster, continuous, and generically complete invariant PDD. 

\subsection{Generically complete continuous invariant PDD}
\label{sub:PDD}

This Subsection reviews the recent invariant $\PDD$ \cite{widdowson2022resolving}, which will be converted into a new graph invariant $\DDG$.

\smallskip
\begin{definition}[Pointwise Distance Distribution]
\label{dfn:PDD}
In the notations of Def.~\ref{dfn:periodic_set},
let $S = \Lambda + M$ be  a periodic set with a motif $M = \{p_1, \ldots, p_m \}\subset U$, where $U$ is a unit cell of a lattice $\Lambda$.
Fix a number of neighbors $k\geq 1$.
For each point $p_i$, let $d_{i1}\leq\cdots\leq d_{ik}$ be a row of Euclidean distances from $p_i$ to its $k$ nearest neighbors in the infinite set $S$.
Consider the matrix $m\times k$ of $m$ distance rows, one for each $p_i\in M$.
If the matrix contains $l\geq 1$ identical rows, collapse them into a single one with the weight $\frac{l}{m}$.
The resulting matrix can be considered as a weighted distribution of rows and is called the \emph{Pointwise Distance Distribution} $\PDD(S;k)$.
\end{definition}

\smallskip
The Pair Distribution Function (PDF) was traditionally used for comparing experimental crystals, but does not distinguish homometric structures \cite{patterson1939homometric}.
Often this PDF is additionally smoothed to guarantee continuity under perturbations but even the exact discrete version of PDF is strictly weaker than the PDD, see details below Fig.~6 in Section 3 of \cite{widdowson2022resolving}. 
The Earth Mover's Distance (EMD) \cite{hargreaves2020earth} defines a continuous metric on PDDs even if they have a different number of rows.
Here is a summary of the PDD advantages over past descriptors.

\begin{enumerate}
\item
$\PDD(S;k)$ is invariant under changes of a unit cell (even for a non-primitive cell) and any Euclidean isometry from the group $E(3)$, see Theorem~3.2 of \cite{widdowson2022resolving}, so $\PDD$ never has \emph{false negatives} for all structures.

\item 
The number $k$ of neighbors can be considered a degree of approximation because increasing $k$ only adds longer distances in extra columns of $\PDD(S;k)$ without changing all shorter distances.
If $k\to+\infty$, the distances from the $k$-th column 
approach $c(S)\sqrt[3]{k}$, where $c(S)$ is inversely proportional to the density of $S$, see Theorem~13 of \cite{widdowson2022average}.

\item $\PDD(S;k)$ can be computed in a near-linear time in both $k$ and the motif size $m$, see Theorem 5.1 of \cite{widdowson2022resolving}, due to a fast nearest neighbor search \cite{elkin2023new}.

\item 
Any periodic point set $S=M+\Lambda$ in general position is \emph{uniquely reconstructable} (up to isometry) using the motif size $m$, a lattice $\Lambda$, and $\PDD(S;k)$ whose largest distance in every row is at least $2R(\Lambda)$, where 
$R(\Lambda)$ is the maximum distance from any point $p\in\mathbb{R}^n$ to $\Lambda$, see Theorem 4.4 of \cite{widdowson2022resolving}.

\item 
$\PDD(S;100)$ distinguished all (more than 670 thousand) periodic crystals in the Cambridge Structural Database (CSD) through more than 200 billion pairwise comparisons over two days on a modest desktop, see Section~6 \cite{widdowson2022resolving}.
Hence all real periodic materials have uniquely defined locations in a common \emph{Crystal Isometry Space} continuously parameterized by the complete invariant isosets \cite{anosova2021isometry}, though the faster $\PDD$ suffices in practice.
\end{enumerate}

The work on Matformer \cite{matformer} notes a potential ambiguity when a point $p\in M$ has different neighbors in $S$ at the same distance.
This ambiguity arises if directions (or vectors) to neighbors are used, while the $\PDD$ takes only distances, which change continuously even if neighbors of $p$ are swapped, see the continuity in Theorem~4.3 of \cite{widdowson2022resolving}.

When writing $\PDD(S;k)$, we put weights into the extra first column.
For storage convenience, rows of $\PDD(S;k)$ can be ordered lexicographically.
Def.~\ref{dfn:PDD} applies to any periodic set, where points can be specific (not all) atoms or centers of molecules in a crystal.

The rock salt crystal NaCl (seen at the top of Fig.~\ref{fig:LuSi})  has a distance $2.81\angstrom$ between adjacent atoms, so $\PDD(\mathrm{NaCl};6)$ is the single row $(1\mid 2.81 \; 2.81 \; 2.81 \; 2.81 \; 2.81 \; 2.81)$, whose first entry is weight 1.
The LuSi crystal at the bottom of Fig.~\ref{fig:LuSi}(a) has four atoms in a unit cell but they collapse into two rows that have a weight of 0.5 and correspond to the atomic types of Lu and Si in the final matrix
$\PDD(\mathrm{LuSi};6)=\left(\begin{array}{c|cccccc}
0.5 & 2.48 & 2.48 & 2.88 & 2.88 & 2.88 & 2.88 \\
0.5 & 2.88 & 2.88 & 2.88 & 2.88 & 2.90 & 3.07 
\end{array}\right)$.
\smallskip

The new graph in Def.~\ref{dfn:DDG} based on the $\PDD$ will include atomic types.
However, when comparing $\PDD$ even without atomic types for all periodic crystals in the CSD, Section 7 of \cite{widdowson2022average} reported five pairs of duplicates, where all geometric parameters were identical to the last digit but one atom was replaced with a different one, for example, Cd with Mn in the pair HIFCAB vs JEPLIA. Indeed, replacing one atom with a larger one should perturb distances to neighbors at least slightly. This has led to five journals investigating the data integrity of the relevant publications.
\smallskip
\begin{definition}[Distance Distribution Graph $\DDG$]
\label{dfn:DDG}
Let $S = M+\Lambda\subset\mathbb{R}^n$ be a periodic point set with a lattice $\Lambda$ and a motif $M = \{p_1, p_2, \ldots, p_m\}$ of points with atomic numbers $t(p_i)$, $i=1,\dots,m$.
Then any point $q\in S$ is obtained by a lattice translation from a point $p_i\in M$ and has the atomic number $t(q)=t(p_i)$.
Fix a number of neighbors $k \geq 1$.
For any point $p\in S$, let $\mathbf{d}(q) \in \mathbb{R}^k$ be the vector of distances in non-decreasing order to its $k$-nearest neighbors in $S$, and $\mathbf{t}(q) \in \mathbb{N}^k$ be the vector of their respective atomic numbers. 
We define an \emph{equivalence relation} $\sim$ on $S$ so that the equivalence class $[p]$ of $p\in S$ consists of all points $q\sim p$ with \textsc{(i)}\label{dfn:ddg_cond1} $t(p) = t(q)$, \textsc{(ii)}\label{dfn:ddg_cond2} $\mathbf{d}(p) = \mathbf{d}(q)$, and \textsc{(iii)}\label{dfn:ddg_cond3} $\mathbf{t}(p) = \mathbf{t}(q)$.
For any point $c\in S$, we define a partial order on all its neighbors in the set $S$ by saying that $p<q$ if one of the following four ordered inequalities holds strictly while all previous ones should be equalities: $|p-c|\leq|q-c|$, $t(p) \leq t(q)$, $\mathbf{d}(p) \leq \mathbf{d}(q)$,  $\mathbf{t}(p) \leq \mathbf{t}(q)$, where the last two inequalities between vectors are lexicographic.
If all inequalities are equal, the neighbors $p,q$ are equivalent: $[p]=[q]$. The \emph{vertices} of the \emph{Distance Distribution Graph} $\DDG(S;k)$ are the equivalence classes $[p]$ for all $p\in S$ (or $M$), which is not larger than $M$ since $[p]$ includes all lattice translates $p+\Lambda$ for any $p\in S$.
The \emph{weight} of a vertex $[p]$ is $|[p]\cap M|/m$, where 
$|[p]\cap M|$ is the number of points $q\in M$ equivalent to $p$.  
Any vertex $[p]$ for $p\in M$ has $k$ \emph{directed edges} $([p],[q])$  in $\DDG(S;k)$ for the first $k$ nearest neighbors $q\in S$ of $p$, ordered as explained above.
\end{definition}

\smallskip

In the rock-salt crystal NaCl, all atoms split into two classes [Na] and [Cl].
For any atom (say, Na), its first 6 nearest neighbors (Cl) are equivalent.
Hence $\DDG($NaCl$;6)$ has two vertices [Na] and [Cl] connected by 6 edges ([Na],[Cl]) and 6 edges ([Cl],[Na]). 

\subsection{Crystal Graph }
\label{sub:GNN}

The conversion of the PDD to the new graph invariant $\DDG$ loses the continuity property because (say) doubling a unit cell and perturbing a motif doubles the vertex set of $\DDG$.
Since the PDD is ultra-fast and generically complete, even this degradation keeps it competitive with past graph invariants that are all discontinuous for the same reason and also need extra parameters such as a cut-off radius making them incomplete by construction.
Def.~\ref{dfn:CG} formalizes the widely used Crystal Graph, which was informally introduced after Fig.~1 in \cite{CGCNN}.

\smallskip
\begin{definition}[Crystal Graph]
\label{dfn:CG}
In the notations of Def.~\ref{dfn:periodic_set},
let $S = M+\Lambda$ be  a periodic set with a motif $M$
and  a lattice $\Lambda$.
Fix a number of neighbors $k\geq 1$ and a \emph{cut-off radius} $r>0$. 
For any point $p \in M$, let $\mathcal{N}(p;k,r)$ be a set of at most $k$ nearest neighbors of $p$ in $S$ within the cut-off radius $r$.
The \emph{Crystal Graph} $\CG(S;k,r)$ has the vertex set $M$ and a directed edge $(p,(q+\Lambda)\cap M)$ for any $p\in M$ and $q\in \mathcal{N}(p;k,r)$, where 
$(q+\Lambda)\cap M$ is the point in $M$ that equivalent to $p$ under a lattice translation.
\end{definition}
\noindent
The Crystal Graph can have up to $k$ multiple edges between the same vertices (atoms in a unit cell), see Fig. ~\ref{fig:LuSi}.
$\CG(S;k,r)$ is invariant of $S$ only if we choose a primitive unit cell because doubling a cell doubles the vertex set.
Since many real crystals are highly symmetric, the vertex set of only non-isometric atoms in $\DDG(S;k)$ is often much smaller than the motif of $S$.
\smallskip

There exists a relationship between the crystal graph and DDG of a crystal structure. If the resulting crystal graph contains an outdegree of $k$ for each node in the graph, it can be converted into a DDG by identifying vertices with the same topology. For vertices $p,q \in M$, the topology of the two vertices is considered equivalent if the outgoing edges of each satisfies $\mathbf{d}(q) = \mathbf{d}(p)$ and $\mathbf{t}(q) = \mathbf{t}(p)$. That is, two vertices have the same edge weights and are directed towards the same type of vertices. In this way, the DDG can result from compressing the crystal graph (should the aforementioned conditions be met).

\subsection{Adaptation for Line Graphs}\label{sec:line_graphs}
In some property prediction models, the inclusion of a \textit{line graph} \cite{Harary1960_line_graph_og,10.1007/BFb0067366_line_graph_def} is used to add additional geometric information such as bond and dihedral angles \cite{ALIGNN,matformer}. The line graph $L(G)$ of a graph $G$ can be defined like so:

\smallskip
\begin{definition}[Line Graph]\label{def:line_graph}
The line graph $L(G) = (V_{L}, E_L)$ of the graph $G = (V, E)$ where $V = \{ v_1, \ldots, v_n \} $ is the set of vertices and $E = \{ e_1,\ldots, e_m \}$ is the set of edges, is defined by the set of vertices $V_{L} = E$ and the edge set $E_{L} = \{ (e_i, e_j): \exists v_k \in e_i \cap e_j  \}$. In the directed case, there exists an edge in the corresponding line graph if for two line graph vertices $e_i = (v_a, v_b) $ and $ e_j = (v_c, v_d)$, $b = c$.
\end{definition}
\smallskip

While the line graph can be incorporated to improve the performance of graph learning algorithms, it can also result in a significant increase in computational load. The number of vertices and edges in the resulting line graph can vastly outnumber the number in the original graph. The order of $L(G)$ is equivalent to the cardinality of the original graph's edge set $E$. The degree of a given vertex $e_i = (v_i, v_k)$ in $L(G)$ is defined by  $|\mathcal{N}(e_i)| = |\mathcal{N}(v_i)| + |\mathcal{N}(v_j)| - 2$ \cite{10.1007/BFb0067366_line_graph_def} in the undirected case. Decreasing the size of the original graph is therefore largely beneficial as such a reduction will be propagated to the line graph.

If a DDG is created according to Def. \ref{dfn:DDG}, there is a possibility that two motif points will be grouped that have different topologies in the line graph resulting from Def \ref{def:line_graph}. To prevent this loss of information, we introduce and use two stronger notions of equivalence on the periodic set $S$. The first notion we call strong equivalence. Informally, we say that two points $p$ and $q$ are strongly equivalent if they are equivalent (according to Def.4), the distance between the sets of their $k$-nearest neighbors is 0, and for every $i \in [k]$ the distances between the set of $k$-nearest neighbors of the $i^{th}$ neighbor of $p$ in $S$ and the set of $k$-nearest neighbors of the $i^{th}$ neighbor of $q$ in $S$ are all equal to 0. The formal definition of this equivalence notion is given below in \textsc{(iv}.a\textsc{)}, and we formally define the used notion of distance between finite point sets in the next paragraph. The second notion, which we call atomic strong equivalence, is defined similarly taking into account atomic types, and is formally defined below in \textsc{(iv}.b\textsc{)}.


The PDD is invariant and establishes a continuous metric via the EMD on periodic sets, but this also applies to simpler finite point sets. Note that when taking the PDD of a finite point set, the parameter $k$ is implicitly determined to be equal to one less than the cardinality of the point set. Let $P$ and $Q$ be the point sets for the $k$-nearest neighbors of $p$ and $q$ (including both $p$ and $q$), respectively. The PDD of $P$ (and $Q$) is the real-valued matrix $\text{PDD}(P; k) \in \R^{(k+1) \times (k+1)}$. Each point set contains $k+1$ points and the weights in the first column cause the PDD to have $k+1$ columns. To compare the PDDs, each row of ordered distances is compared with each row in the opposing PDD (without considering the weights in the first column). The result is a distance matrix where the entry in the $i^{th}$ row and $j^{th}$ column is the cost for transporting from the $i^{th}$ point in $P$ to the $j^{th}$ point in $Q$ in the subsequent minimum-cost flow problem. The weights in the first column of $\text{PDD}(P)$ and $\text{PDD}(Q)$ form the distributions to be transported. The cost for the solution to this problem is the distance between $P$ and $Q$, also referred to as the Earth Mover's Distance. If $P$ and $Q$ are found to be equivalent (having a distance of zero), the point sets of their neighbors are then also compared using the same technique. Should each pair of $k$ neighbors also be equivalent, the motif points $p$ and $q$ are considered part of the same equivalence class. We refer to this as condition \textsc{(iv}.a\textsc{)}. Formally,

\begin{itemize}
    \item[(\textsc{iv}.a)] Two points $p$ and $q$ in the motif $M$ of a periodic set $S = M + \Lambda \subset \R^n$ are members of the same equivalence class $[p]$ if they satisfy conditions \textsc{(i-iii)} in Def. \ref{dfn:DDG}, as well as $\text{EMD}(\text{PDD}(N(S; k; p), \text{PDD}(N(S; k; q)) = 0$, and $\text{EMD}(\text{PDD}(N(S; k; p_i), \text{PDD}(N(S; k; q_i)) = 0$ where $p_1 < p_2 \ldots < p_k$ and $q_1 < q_2 \ldots < q_k$ are ordered according to Def \ref{dfn:DDG}.
\end{itemize}

While condition \textsc{(iv}.a\textsc{)} establishes the distance between finite point sets, it considers these points unlabeled and thus, does not take into consideration the type of point. In the case of materials, each point has an atomic species that should be accounted for. To do this we use the \textit{Atomic Mass Weighted PDD}.

\smallskip
\begin{definition}[Atomic Mass Weighted Pointwise Distance Distribution]\label{def:mPDD}
    For a labeled periodic point set $S = M + \Lambda \in R^n$ where each point $p_i \in S$ carries a label equivalent to its corresponding atomic mass $a(p_i)$, the Atomic Mass Weighted PDD of $S$, \emph{mPDD$(S;k)$}, is equivalent to the PDD of $S$ where rows are not grouped, having each row's final weight, $w_i$, defined by $w_i = a(p_i) / \sum_{j=1}^{|M|} a(p_j)$. 
\end{definition}
\smallskip
\noindent
In this version of the PDD, rows are not grouped together as they are in the original. This is done to prevent losing atomic information in the case when two motif points have the same $k$-nearest neighbor distances, but correspond to different atomic types.

The same procedure mentioned above for determining strong equivalence is followed but using the mPDD. We refer to this as condition \textsc{(iv}.b\textsc{)}. Formally,

\begin{itemize}
    \item[(\textsc{iv}.b)] Two points $p$ and $q$ in the motif $M$ of a periodic set $S = M + \Lambda \subset \R^n$ are members of the same equivalence class $[p]$ if they satisfy conditions \textsc{(i-iii)} in Def. \ref{dfn:DDG}, as well as $\text{EMD}(\text{mPDD}(N(S; k; p), \text{mPDD}(N(S; k; q)) = 0$, and $\text{EMD}(\text{mPDD}(N(S; k; p_i), \text{mPDD}(N(S; k; q_i)) = 0$ where $p_1 < p_2 \ldots < p_k$ and $q_1 < q_2 \ldots < q_k$ are ordered according to Def \ref{dfn:DDG}.
\end{itemize}

Once either of the aforementioned conditions is applied and the equivalence classes are created the construction of the DDG proceeds as normal. Def. \ref{def:line_graph} is followed to produce the line graph. Together with the original DDG, we refer to the line graph and any line graph derived thereof generally as \textit{distribution graphs}.

\subsection{Crystal Graph Convolutional Neural Network}

Below we modify traditional graph neural networks to take advantage of weighted vertices in $\DDG(S;k)$. 
Changes must be made to two specific operations: normalization and graph readout. The computations we propose here can be applied in a general setting for GNNs which take advantage of a number of different mechanisms including convolution, attention, or message-passing layers. For consistency, we will use CGCNN which was proposed in the original paper from which the Crystal Graph was created, though the DDG is not limited to this model. 
\smallskip

A single convolution from CGCNN can be defined by the equation \cite{CGCNN},
\begin{equation}
    \mathbf{v}_i^{(t+1)} = \mathbf{v}_i^{(t)} + \sum_{j,k}\sigma (\mathbf{z}_{(i,j)_k}^{(t)} \mathbf{W}_f^{(t)} + \mathbf{b}_f^{(t)})\odot  g(\mathbf{z}_{(i,j)_k}^{(t)} \mathbf{W}_s^{(t)} + \mathbf{b}_s^{(t)})
\end{equation}
\noindent
where
\begin{equation*}
    \mathbf{z}_{(i,j)_k}^{(t)} = \mathbf{v}_i^{(t)} \oplus \mathbf{v}_j^{(t)}  \oplus \mathbf{u}_{(i,j)_k}
\end{equation*}
and $\mathbf{v}_i^{(t)}$ is the embedding of the $i^{th}$ vertex after $t$ convolutions, $\mathbf{u}_{(i,j)_k} $ are the edge features of $k^{th}$ edge between vertices $i$ and $j$ and $\mathbf{W}_f^{(t)}, \mathbf{W}_s^{(t)}$ and $ \mathbf{b}_f^{(t)},  \mathbf{b}_s^{(t)}$ are the learned weight and bias matrices, respectively. The operators $\oplus$ and $\odot$ refer to concatenation and element-wise multiplication, respectively.

\subsection{Atomistic Line Graph Neural Network}

ALIGNN incorporates the angles of atom triplets by using a line graph to update the edges of the original crystal graph. The vertex embeddings are updated according to
\begin{equation}\label{eq:node_update}
    \mathbf{v}_i^{(t+1)} = \mathbf{v}_i^{(t)} + \text{SiLU} \bigg( \text{BN}\bigg( \mathbf{W}_s^{(t)} \mathbf{v}_i^{(t)} + \sum_j \mathbf{u}_{i,j}^{(t)} \mathbf{W}_d^{(t)} \mathbf{v}_j^{(t)} \bigg) \bigg)
\end{equation}
where $\mathbf{v}_i^{(t)}$ is the embedding of the $i^{th}$ vertex after $t$ convolutions, $\mathbf{u}_{(i,j)} $ are the edge features of the edge between vertices $i$ and $j$ and $\mathbf{W}_s^{(t)} $ and $ \mathbf{W}_d^{(t)}$ are the learned weight matrices for the source and destination vertices; BN is batch normalization and SiLU is the Sigmoid Linear Unit \cite{hendrycks2016gaussian}. The edge features of the graph are updated using the following:
\begin{equation}\label{eq:edge_update}
    \mathbf{u}_{i,j}^{(t+1)}  = \mathbf{u}_{i,j}^{(t)} + \text{SiLU} \big( \text{BN}\big( \mathbf{W}_a^{(t)} \mathbf{v}_i^{(t-1)} + \mathbf{W}_b^{(t)} \mathbf{v}_j^{(t-1)} + \mathbf{W}_c^{(t)} \mathbf{u}_{i,j}^{(t-1)} \big) \big)
\end{equation}
Eq. \ref{eq:node_update} and \ref{eq:edge_update} are used to update both the original and line graphs. The resulting updated vertices of the line graph are used in the update for the original graph's edges. 

\subsection{Including PDD Weights}

Normalization is a data scaling tool commonly used to increase model stability and improve training efficiency \cite{whynorm}. Two commonly used types are batch normalization \cite{batch_norm} and layer normalization \cite{layernorm}; each of which operates on different dimensions of the input. Layer normalization scales features with respect to one another, within a single sample. In batch normalization, normalization is done over multiple samples for a single feature. 
\smallskip

In GNNs, the result of layer normalization on vertices is independent of the number of vertices in any given graph, since the computation operates on the features of one graph. 
The same cannot be said of batch normalization. In a single batch, it is necessary to operate on multiple graphs at once. These graphs have no requirement to consist of the same number of vertices. 
Due to this disparity, a single graph can be over-represented in the mean or variance computed during normalization. 
\smallskip

While it may not be a problem for other GNNs, graphs used to represent crystals that depend on the unit cell can fall victim to this ambiguity. One could arbitrarily scale a crystal's unit cell size, or select different basis vectors and predictions for the model trained on this altered data would be different despite using the same crystals and hyper-parameters. 
The DDG alleviates the concerns over unit cell selection by the use of weights within batch normalization. 
The weighted batch normalization for a batch $b$ containing graphs $G_1, \ldots G_n$ with a respective number of vertices $g_1, \ldots g_n$ can be computed using 

\begin{equation}\label{eq:weight_bn_mean}
    \bm{\mu}_b = \frac{\displaystyle\sum_{j=1}^n \sum_{i=1}^{g_j} w_{ij} \mathbf{v}_{ij}}{B}
\end{equation}
\begin{equation}\label{eq:weight_bn_var}
    \bm{\sigma}_b^2 = \frac{\displaystyle\sum_{j=1}^n \sum_{i=1}^{g_j} w_{ij}(\mathbf{v}_{ij} - \bm{\mu}_b)^2}{B}    
\end{equation}
\begin{equation}\label{eq:normalize}
    N(\mathbf{v}) = \frac{\mathbf{v} - \bm{\mu}_b}{\bm{\sigma}_b}
\end{equation}
where $B = \sum_{j=1}^n \sum_{i=1}^{g_j} w_{ij}$, $\mathbf{v}_{ij}$ is the $i^{th}$ vertex embedding in the $j^{th}$ graph, and $w_{ij}$ is the weight of the $i^{th}$ vertex in the $j^{th}$ graph. Eq. \ref{eq:weight_bn_mean} and Eq. \ref{eq:weight_bn_var} are used to compute the weighted mean and weighted biased variance of the batch respectively. 
Eq. \ref{eq:normalize} provides the formula for normalizing a given sample or vertex $\mathbf{v}$ from batch $b$.

The readout or pooling layer of a GNN is used to condense the information contained within the individual vertex embeddings into a single vector representative of the whole graph. This vector is commonly passed to a multi-layer perceptron for a prediction to be made. 

While there are many options for a pooling layer, they all must retain the property of permutational invariance (the order of the vertex embeddings does not matter). Some of the most common graph readouts are maximum and mean pooling and normalized sum. In the former case, no adjustments need to be made as the maximum of each feature within the embedding does not depend on the number of vertices. For the latter two, this is not true. For mean pooling, we must use the weighted average of the vertex embeddings defined by

\begin{equation}\label{eq:weightsum}
    \mathbf{v_c} = \sum_i w_i \mathbf{v}_i
\end{equation}
where $\mathbf{v}_i$ is the vertex embedding after the predefined number of graph convolutions is complete. For the normalized sum, we can again use the weighted average described by Eq. \ref{eq:weightsum} and then apply layer normalization.

\section{Experiments on Materials Project  and T2 crystals}
\label{sec:experiments}

We test the strength of the DDG on the T2 dataset \cite{Pulido2017-ff} of 5,687 simulated molecular organic crystals and on the Materials Project database of 36,678 crystals \cite{MP1} obtained by extra optimization from real mostly inorganic crystals. The T2 dataset was produced during the crystal structure prediction process using quasi-random sampling \cite{case2016convergence_quasi_random}. 
We focus on property prediction of lattice \cite{latticeenergy,latticeenergy2}, formation \cite{formationenergy}, and band gap energy \cite{bandgap}. 

Each dataset presents a different challenge for the model. The Materials Project crystals are varied in their composition, while all crystals in the T2 dataset have the same molecular composition. For each dataset, the primary differentiator is different, and applying the model to both is important in displaying its flexibility and robustness.  

In the first experiment, we will introduce an additional hyper-parameter for the model and show how it can be used to compress the graph representation. We will then show how this compression affects the accuracy of the model. 
Then we will present a method for finding the optimal $k$ when constructing the DDG. 

The Jarvis-DFT dataset \cite{choudhary2020joint_jarvis_og}  is introduced and a modified version of ALIGNN which uses the DDG is tested against the original model \cite{ALIGNN}.

Finally, we show the importance of having a graph that is independent of the unit cell.

\subsection{Training \& Evaluation}
All experiments will be done using 80/10/10 training/validation/test splits. The atom embeddings used will be identical to those used in CGCNN \cite{CGCNN}. The various atomic properties and their dimensionality are shown in Table \ref{tab:features}.

\begin{table}[h]
\centering
\caption{\label{tab:features}%
Atom feature embeddings by atomic property. }
\begin{tabular}{p{4cm}c}
\toprule
Atomic Property &
Dimensionality\\
\hline
Group Number & 18\\
Period & 9\\
Electronegativity & 10\\
Covalent Radius & 10\\
Valence Electrons & 12\\
First Ionization Energy & 10\\ 
Electron Affinity & 10\\
Block & 4\\
Atomic Volume & 10\\
\bottomrule
\end{tabular}
\end{table}

The edge embeddings will also be identical to that of CGCNN. Embeddings will consist of a one-hot encoding of the edge weight (the distance between atoms in angstroms) using Gaussian smoothing.

Once the model is trained, MAE will be used to present results. MAE is an error measure defined by \cite{mae}:
$$ \text{MAE} = \frac{\sum_{i=1}^{n} \lvert y_i - x_i \rvert }{ n } $$
where $y_i, x_i$, and $n$ are the true property value, predicted property value, and number of samples, respectively. This measure is used throughout other works in crystal property prediction \cite{ALIGNN, CGCNN, crysXPP, iCGCNN}, thus we selected it to make for easier comparisons.
\smallskip

For CGCNN, all results shown were produced after the model was trained for 200 epochs using ADAM\cite{ADAM} optimizer with mean-squared error loss (except where otherwise stated). Results using stochastic gradient descent were worse overall for both the Crystal Graph and DDG, but the underlying trend in performance remained the same. The model implementation used was that of the original CGCNN paper with the added code to create the DDG and the ability to set the new hyper-parameters. We provide a custom weighted batch normalization layer, while the non-weighted version uses the default implementation in the model.

\subsection{Impact of Collapse Tolerance on Predictions}\label{sec:col_tol_testing}

\begin{center}
\begin{figure}
\centering
\includegraphics[width=\linewidth]{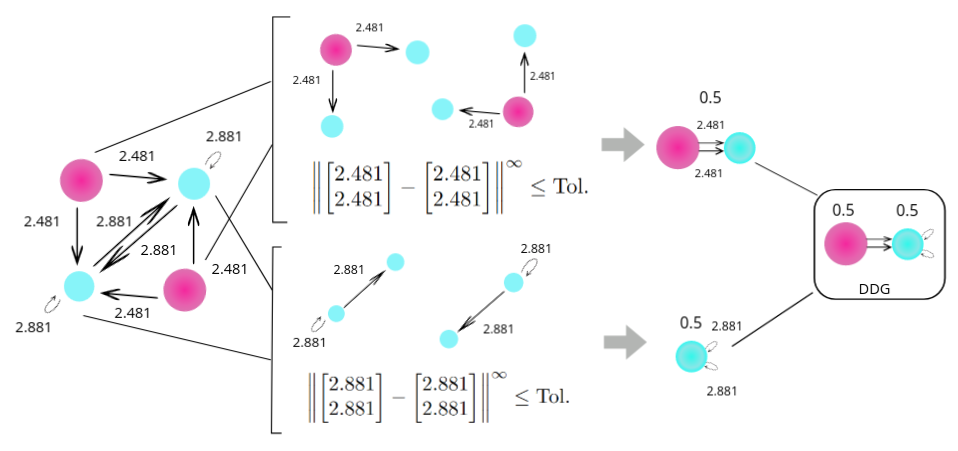}
\caption{\label{fig:col_tol} Transformation from unit cell based graph to the DDG including the collapse tolerance which allows for small differences in the edges features while still grouping similar vertices (atoms). }
\end{figure}    
\end{center}

The grouping of atoms occurs under three conditions, one of which is that the $k$-nearest neighbors' distances must be the same. This presents a problem when creating the implementation for the DDG as small discrepancies can arise through means such as floating point arithmetic, measurement error, or atomic perturbations. For this reason, we introduce a tolerance such that, if the difference between $k$-nearest neighbors' distances is less than this tolerance, they will be considered the same. We will refer to this tolerance as the \textit{collapse tolerance} and it will be considered a hyper-parameter. Fig. \ref{fig:col_tol} displays the application of the collapse tolerance on Lutetium Silicon. Note that the comparisons of the pairwise Euclidean distances are conducted on atoms that have the same atomic species, one of the requirements mentioned in Def. \ref{dfn:DDG}.

In this experiment, we will look at how the collapse tolerance affects the accuracy of model predictions. Intuitively, the collapse tolerance presents a trade-off. As it increases, atoms will be grouped, creating a more compressed representation. However, when the tolerance reaches a point at which it is large enough to group atoms that should be treated separately, information loss will begin to occur and this should have a negative impact on prediction accuracy. 
\smallskip

Table \ref{tab:bycoltolmp} shows the resulting mean absolute errors (MAE) across each of the collapse tolerances. The hyper-parameter $k$ for both graphs will be the same. The convolutional layer developed by the authors of CGCNN can appropriately weigh the influence of distant atoms accordingly, so their interactions are weakened. As such, when testing the Crystal Graph we use an adequately large $r$ to have $k$ neighbors for each vertex.
\smallskip

Across both properties, the DDG is able to beat the prediction accuracy of the Crystal Graph while remaining significantly smaller with respect to the number of vertices and edges. As the tolerance increases the graph is reduced in size. There is a threshold where performance should begin to suffer due to information loss. We can see this begin to happen, though it does not manifest in large increases in error. Overall, the differences in the results across the tolerances are relatively small.

\begin{table}[h]
\centering
\caption{\label{tab:bycoltolmp}%
The Mean Absolute Error (MAE) for predicting the formation energy (FE in eV) and band gap energy (BG in eV/atom) for Materials Project crystals and the DDG with several collapse tolerances and the Crystal Graph (CG) for $k=12$ nearest neighbors. 
The relative size is the average percent of the number of vertices compared to the CG.}
\begin{tabular}{ccccc}
\toprule
Graph&
Tolerance &
Band Gap  MAE $(eV)$&
Formation Energy MAE $(eV/atom)$&
Relative Size\\ 
\hline
CG & N/A & 0.327 & 0.0512 & 100.0\% \\
DDG & $10^{-8}$ & \textbf{0.314} & \textbf{0.0456} & 48.8\% \\
DDG & $10^{-6}$ & 0.315  & 0.0468 & 47.5\% \\
DDG & $10^{-4}$ & 0.316 & 0.0469 & 43.7\% \\
\bottomrule
\end{tabular}
\end{table}

Work from \textit{DeeperGATGNN} \cite{deeperGNN} showed that convergence for CGCNN can come after training for a number of epochs closer to $500$. We replicate the same experiment done in Table \ref{tab:bycoltolmp} after increasing the number of epochs to 500 at a collapse tolerance of $10^{-4}$ in Table \ref{tab:bycoltolmp500}.

\begin{table}[h]
\centering
\caption{\label{tab:bycoltolmp500}%
The Mean Absolute Error (MAE) for predicting the formation energy (FE in eV) and band gap energy (BG in eV/atom) for Materials Project crystals and the DDG with a collapse tolerance of $10^{-4}$ and the Crystal Graph (CG) for $k=12$ nearest neighbors at 200 and 500 epochs. The relative size is the average percent of the number of vertices compared to the CG.}
\begin{tabular}{cccccc}
\toprule
Graph&
Tolerance & 
Epochs & 
BG  MAE $(eV)$&
FE MAE $(eV/atom)$&
Relative Size\\ 
\hline
CG & N/A & 200 & 0.327 & 0.0512 & 100.0\% \\
DDG & $10^{-4}$ & 200 & 0.316 & \textbf{0.0469} & 43.7\% \\
\hline
CG & N/A & 500 & 0.328 & 0.0478 & 100.0\% \\
DDG & $10^{-4}$ & 500 & 0.319 & \textbf{0.0455} & 43.7\% \\
\bottomrule
\end{tabular}
\end{table}

While the gap between the formation energy MAEs has decreased, the band gap results are marginally different. The DDG provides a decrease of $4.8\%$ at 500 epochs and $8.4\%$ at 200 epochs. For band gap energy this improvement stands at $3.4\%$ and $2.7\%$ for 200 and 500 epochs, respectively.

Table \ref{tab:bycoltolt2} shows the results for the analogous experiment on the T2 dataset for the prediction of lattice energy. For the T2 crystals, we use a more aggressive tolerance of $10^{-3}$ as $10^{-8}$ is inadequate in significantly decreasing the size of the graphs. 

The T2 dataset consists of molecular crystals all based on triptycene. The primary differentiator between these crystals is their geometric arrangement. These structural differences manifest themselves as changes in the pairwise distances between atoms. Such changes can be subtle. Because the collapse tolerance provides a way for atoms with (slightly) different $k$-nearest neighbor distances to be grouped, it is possible that these differences in structure can be lost. 

\begin{table}[h]
\centering
\caption{\label{tab:bycoltolt2}%
The Mean Absolute Error (MAE) for predicting the lattice energy (LE in kJ/mol) for T2 crystals and the DDG with several collapse tolerances and the Crystal Graph for $k=16$ nearest neighbors. 
The relative size is the average percent of the number of vertices compared to the Crystal Graph.}
\begin{tabular}{cccc}
\toprule
Graph&
\multicolumn{1}{c}{\textrm{Tolerance}}&
\multicolumn{1}{c}{\textrm{Lattice Energy MAE $(kJ/mol)$}}&
\multicolumn{1}{c}{\textrm{Relative Size}}\\ 
\hline
CG & N/A & 3.358 & 100.0\% \\ 

DDG & $10^{-6}$ & \textbf{3.154} & 31.4\%\\ 
DDG & $10^{-4}$ & 3.256 & 28.4\%\\ 
DDG & $10^{-3}$ & 3.312 & 27.2\%\\ 
\bottomrule
\end{tabular}

\end{table}

\smallskip

Such information loss should result in a decrease in prediction accuracy. Alternatively, it is possible for differences in structure to be small enough that the property values change insignificantly. In this case, the collapse tolerance is useful for removing such noise. This in turn could allow the model to better generalize and prevent overfitting, ultimately improving results.
\smallskip

In Table \ref{tab:bycoltolt2}, at a tolerance of $10^{-6}$, we can see the grouping of atoms in the representation lowers the overall MAE. This is in line with our previous hypothesis as such a tolerance value is still conservative. When the tolerance increases past this, the MAE increases significantly, approaching that of the Crystal Graph. This is where information loss begins to occur. 

If we consider the best-performing collapse tolerance, we are able to shrink the crystals to just $31.4\%$ of the original size of the Crystal Graph and reduce MAE by $6.1\%$.

\subsection{Selection of $k$-Nearest Neighbors}\label{sec:k_sens_testing}

The value of $k$ is considered a DDG hyper-parameter to tune for best prediction performance. Previous graph representations which also use each atom's nearest neighbors have selected this $k$ based on trial and error, without providing much guidance on how this selection should be conducted. Here we develop a method to help find an adequate $k$ using the properties of the PDD. Selection of $k$ should always be the smallest possible without sacrificing prediction quality.

We know the PDD is generically complete given $\Lambda, m$ and a $k$ large enough that the values in the final column of the PDD are all larger than twice the \emph{covering radius} of the lattice $\Lambda$ for a given periodic set, see Theorem 4.4 of \cite{widdowson2022resolving}. This upper bound for $k$ can be very large but is not without value.

Since we group together atoms that experience the same $k$ nearest neighbor distances (among other conditions), as $k$ increases the number of groups will either stay the same or increase (when the $k^{th}$ distance finally becomes different and grouping can no longer occur). Using this, we can establish a lower bound defined by a large enough $k$ that the groupings are the same as our upper bound. This method establishes a range in which to search for optimal $k$. 

Every crystal in the dataset will have differing ranges, and applying each one individually such that each graph uses its own $k$ value yields poor results. This is not surprising, the Earth Mover's Distance \cite{hargreaves2020earth,rubner2000earth} establishes a continuous metric between PDDs with fixed $k$ according to Theorem 4.3 of \cite{widdowson2022resolving}. So, given a different range for each sample in the data, we want a single value for $k$ that is sufficient across the dataset. 

We take the maximum of this lower bound for the crystals in our dataset, without considering outliers. What is considered an outlier is not well defined, so we will convey this in terms of what percent of the dataset this $k$ value satisfies the aforementioned condition. At $k=6$, the lower bound is satisfied for $98\%$ percent of samples, and at $k=9$, this percentage rises to $99\%$. 

\begin{table}[h]
\centering
\caption{\label{tab:bykMP}%
Prediction MAE using the DDG on formation energy (FE) in eV/atom and band gap (BG) energy in eV when the number of nearest neighbors $k$ is varied at a collapse tolerance of $10^{-4}$. }

\begin{tabular}{cccc}
\toprule
$k$-Nearest Neighbors&
\multicolumn{1}{c}{\textrm{Coverage}}&
\multicolumn{1}{c}{\textrm{Band Gap MAE $(eV)$}}&
\multicolumn{1}{c}{\textrm{Formation Energy MAE $(eV/atom)$}}\\
\hline
3 & 87.9\% & 0.333 & 0.0584\\
6 & 98.7\% & 0.299 & 0.0509\\
9 & 99.1\% & \textbf{0.295} & 0.0477\\
12 & 99.2\% & 0.316 & \textbf{0.0456}\\
15 & 99.3\% & 0.319 & 0.0457\\
18 & 99.4\% & 0.317 & 0.0474\\

\bottomrule
\end{tabular}

\end{table}

Table \ref{tab:bykMP} shows the resulting MAE for the formation and band gap energy after $k$ is varied. At $k=3$ the results deteriorate and the number of neighbors should be increased. Formation energy peaks at $k=12$ and band gap energy at $k=9$.

When the number of neighbors is increased, the degree of the vertices of the graph is increased. By doing this, the information from more distant atoms is included during the convolution. The convolution for a given vertex aggregates information pertaining to the neighboring vertices. 

As $k$ increases, the local information about the atom is traded in exchange for global information since more distant atoms are now included in the convolution and thus, the new vertex embedding. At a point, with $k$ being large enough, this causes over-saturation and the local information of the atom is diluted. In both band gap and formation energy this can be seen in the higher $k$ values when MAE begins to increase after reaching a minimum. Furthermore, we can deduce that if performance peaks at a lower $k$ then the vertex embeddings are optimally produced when using atoms that are closer in distance.

\subsection{Impact of Unit Cell choice}

A valid unit cell $U$ for a given crystal can be transformed by multiplying the original basis vectors $(\mathbf{a}, \mathbf{b}, \mathbf{c})$ by a transformation matrix $\mathbf{P}$ where $P_{ij} \in \mathbb{Z}$ 
\begin{equation*}
    \begin{pmatrix}
P_{11} & P_{12} & P_{13}\\
P_{21} & P_{22} & P_{23}\\
P_{31} & P_{32} & P_{33}
\end{pmatrix}
\end{equation*}
such that $det(\mathbf{P})  = 1$ to maintain cell volume and $det(\mathbf{P}) > 1$ for a supercell \cite{arnold2006transformations}, a unit cell which has a larger volume than the original. 

\begin{center}
\begin{figure}
\centering
\includegraphics[width=0.7\linewidth]{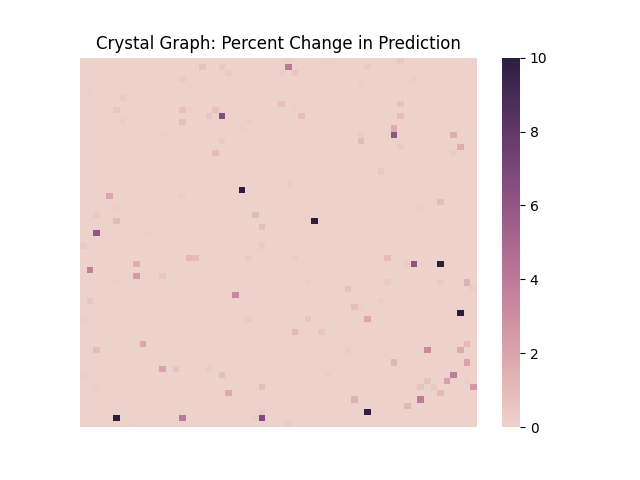}
\caption{\label{fig:supercell_cg} Absolute Percent difference in prediction of formation energies for the test set using the Crystal Graph after altering samples in the training data to have supercells. }
\end{figure}    
\end{center}

In this experiment, these transformations are done to the unit cell to form a supercell for $2,000$ of the crystals (roughly $7\%$) from the training set of the Materials Project data. The amount was chosen to show that even just a small fraction of crystals with a changed unit cell can affect the prediction results of the whole dataset. A control run was done using the original data, and then the unit cells were altered for a second run. We chose to use formation energy as the band gap energy data has very small values that can result in inflated percent change. 

A heatmap of the absolute percent change in prediction between the first and second run for the Crystal Graph is shown in Figure \ref{fig:supercell_cg}. On average the predictions changed by $4.36\%$, with the maximum change being over $135\%$ compared to its original value. The overall test MAE increased slightly from $0.501$ to $0.502$. 

With the DDG, the resulting graph would be the exact same regardless of the transformation to the unit cell. Intuitively, this can be seen as the PDD keeps track of the proportion of a particular atom within the cell through its weights and eliminates atoms with the same behavior through collapsing rows in the PDD. Such collapsing removes the dependence of our graph from the unit cell choice. In order to maintain such consistency using the Crystal Graph, users would need to apply the same cell reduction algorithm (for example) to obtain a particular cell, in line with what others have used. Even this is not without problems, as reduction techniques are discontinuous under atomic perturbations or errors in experimental measurement \cite{perturbations}.

\subsection{Application on Line Graphs}

We apply ALIGNN to the Jarvis-DFT dataset \cite{choudhary2020joint_jarvis_og}. The properties included aim to present a variety of sample sizes to show their effect on the performance of the Crystal Graph against the Distribution Graph. The implementation of the crystal graph for ALIGNN differs slightly from its implementation in the original work. Specifically, the resulting graph produced by ALIGNN's implementation has exactly $k$ neighbors for each motif point included in the graph. Further, backward edges are placed between all pairs of vertices, effectively doubling the number of edges in the representation. This increase is then propagated to the derived line graph. To reduce computational cost, we leave out the backward edges added to the original graph, but these can be accounted for in the distribution graph by simply doubling $k$.

Each version of ALIGNN is trained for $300$ epochs (the same as the original paper \cite{ALIGNN}) using AdamW \cite{loshchilov2018decoupled_adamw} optimizer and a batch size of $64$. The model parameters remain the same as in the original implementation except for the hidden embedding size which was reduced from $256$ to $128$. This was done to reduce memory requirements and ensure training and validation errors converged.

\begin{table}[h]
    \centering
        \caption{ Prediction MAE of ALIGNN on the Jarvis-DFT dataset using the Crystal Graph (CG) and the Distribution Graph (DG) created with condition \textsc{(iv}.a\textsc{)} using $k=12$ and no backward edges. The relative size of the original graph $G$ and the line graph $L(G)$ are calculated by taking the mean ratio of the number of vertices $V$ or edges $E$ in the Distribution graph to the Crystal Graph at a collapse tolerance of $10^{-4}$. Abbreviations of the properties are as follows:  Formation Energy (FE), Shear Modulus (SM), Bulk Modulus (BM), Maximum Piezo stress coefficient ($e_{ij}$ and $d_{ij}$), and Exfoliation Energy (EE).   }
    \begin{small}
  \begin{tabular}{|c|c|c|c|c|c|c|}
    \hline
    \multirow{2}{*}{\hfil Property (units)} &
    \multirow{2}{*}{\hfil $n$ } &
    \multirow{2}{*}{\hfil CG } &
    \multirow{2}{*}{\hfil DG } &
      \multicolumn{3}{c|}{Relative Size of DG to CG } \\
    & & & & \hfil $|V(G)|$ & \hfil $|V(L(G))|$ & \hfil $|E(L(G))|$   \\
    \hline
    FE \hspace{0.1cm} \small{$(eV)/atom$}   & 75.993 & 0.042 & \textbf{0.037} & \hfil 69.4\% & \hfil 66.8\% & \hfil 64.2\%  \\
    SM \hspace{0.1cm} \small{$log_{10}(GPa)$} & 23,824 & 14.41 & \textbf{14.32} & \hfil 72.0\% &  \hfil 69.0\% &  \hfil 66.1\% \\
    BM \hspace{0.1cm} \small{$log_{10}(GPa)$} & 23,824 & 12.92 & \textbf{11.89} & \hfil 71.1\% &  \hfil 68.2\% &  \hfil 65.5\% \\
    $e_{ij}$ \hspace{0.1cm} \small{$Cm^{-2}$}  & 4,799 & 0.122 & \textbf{0.117} & \hfil 64.4\% &  \hfil 63.0\% & \hfil 61.4\% \\
    $d_{ij}$ \hspace{0.1cm} \small{$CN^{-1}$}  & 3,347 & 12.09 & \textbf{11.26}  & \hfil 65.2\% &  \hfil 63.6\% & \hfil 61.9\% \\
    EE \hspace{0.1cm} \small{$meV/atom$}  & 813 & 40.02 & \textbf{36.43} & \hfil 53.5\% & \hfil 52.7\% & \hfil 51.5\% \\
    \hline
  \end{tabular}
    \end{small} 
    \label{tab:jarvis_results}
\end{table}

The prediction MAE results on the test set for ALIGNN are listed in Table \ref{tab:jarvis_results}. For these results, we apply condition \textsc{(iv}.a\textsc{)} which is more lenient than condition \textsc{(iv}.b\textsc{)}. The result is a smaller graph representation at the cost of potential information loss. For properties with a smaller number of samples, this provides a regularization effect. When compared to the results in Table \ref{tab:jarvis_results2}, the MAE is lower for exfoliation error, $e_{ij}$, and $d_{ij}$ despite the graph being smaller in terms of vertices and edges. It is at larger sample sizes that we see performance start to decline, likely because more information is needed to distinguish structures. Regardless, the distribution graph created using condition \textsc{(iv}.a\textsc{)} decreases formation energy MAE by $12\%$, shear modulus by $0.6\%$, bulk modulus by $8\%$, $e_{ij}$ by $4.1\%$, $d_{ij}$ by $6.9\%$ and exfoliation energy by $9\%$. Even in the most conservative improvement, the number of vertices and edges in that graph and line graph are decreased by $28\%$. The distribution graph produced by condition \textsc{(iv}.b\textsc{)} provides a more modest improvement on graph size reduction of at least $11.2\%$. Meanwhile, the decrease in MAE on properties with smaller sample sizes (exfoliation energy, $e_{ij}$, and $d_{ij}$) is less significant at $8.4\%, 0.6\%$ and $2.5\%$. For formation energy, shear modulus, and bulk modulus these values are $16.7\%, 2.2\%$, and $9.3\%$.

\begin{table}[h]
    \centering
        \caption{ Prediction MAE of ALIGNN on the Jarvis-DFT dataset using the Crystal Graph (CG) and the Distribution Graph (DG) created with condition \textsc{(iv}.b\textsc{)} using $k=12$ and no backward edges. The relative size of the original graph $G$ and the line graph $L(G)$ are calculated by taking the mean ratio of the number of vertices $V$ or edges $E$ in the Distribution graph to the Crystal Graph at a collapse tolerance of $10^{-4}$. Abbreviations of the properties are as follows:  Formation Energy (FE), Shear Modulus (SM), Bulk Modulus (BM), Maximum Piezo stress coefficient ($e_{ij}$ and $d_{ij}$), and Exfoliation Energy (EE).   }
    \begin{small}
  \begin{tabular}{|c|c|c|c|c|c|c|}
    \hline
    \multirow{2}{*}{\hfil Property (units)} &
    \multirow{2}{*}{\hfil $n$ } &
    \multirow{2}{*}{\hfil CG } &
    \multirow{2}{*}{\hfil DG } &
      \multicolumn{3}{c|}{Relative Size of DG to CG } \\
    & & & & \hfil $|V(G)|$ & \hfil $|V(L(G))|$ & \hfil $|E(L(G))|$    \\
    \hline
    FE \hspace{0.1cm} \small{$(eV)/atom$}   & 75.993 & 0.042 & \textbf{0.035} &  86.9\% & 84.1\% & 82.9\%   \\
    SM \hspace{0.1cm} \small{$log_{10}(GPa)$} & 23,824 & 14.41 & \textbf{14.10} & 88.1\% & 84.9\% & 83.1\% \\
    BM \hspace{0.1cm} \small{$log_{10}(GPa)$} & 23,824 & 12.92 & \textbf{11.72} &  88.2\% & 85.2\% & 83.5\% \\
    $e_{ij}$ \hspace{0.1cm} \small{$Cm^{-2}$}  & 4,799 & 0.122 & \textbf{0.121} &  88.6\% & 87.1\% & 86.3\% \\
    $d_{ij}$ \hspace{0.1cm} \small{$CN^{-1}$}  & 3,347 & 12.09 & \textbf{11.79}  & 88.8\% & 87.1\% & 86.2\% \\
    EE \hspace{0.1cm} \small{$meV/atom$}  & 813 & 40.02 & \textbf{36.65} & 83.3\% & 82.4\% & 81.8\% \\
    \hline
  \end{tabular}
    \end{small} 
    \label{tab:jarvis_results2}
\end{table}

\section{Discussion}

The methods presented in Subsection \ref{sec:line_graphs} are generally applicable for graphs used in GNNs, however, their application to periodic materials is particularly effective since, by their nature, these crystals exhibit repetitive behavior. For objects that can be represented as point sets (crystals, proteins, molecules, etc.), graphs are often constructed with edges being placed between a given point and its $k$-nearest neighbors. This consistency in outdegree for each vertex makes the application of distribution graphs more effective since it is more likely to find repetitive topologies that can be grouped.

When graphs are fed into a GNN model, the vertex embeddings are typically dependent on the adjacent edge features and neighboring vertex features. The process of converting a given graph to a distribution graph maintains these relationships for any given vertex in the graph and thus, the resulting updated embeddings will be the same. This is conditioned on the collapse tolerance used being exactly zero. For any collapse tolerance greater than zero, the resulting edge embeddings will differ slightly as vertices with slightly different edge features will be averaged.  

The improvement coming from the distribution graph stems from two possible sources: the use of weights instead of multiplicities in the graph and the regularization that is produced by using a positive collapse tolerance. 

The use of weights accomplishes two things: it describes the vertices of the graph in terms of a concentration instead of using multiplicities and it prevents graphs with a larger number of vertices from over-influencing the mean and standard deviation during batch normalization. In the former case, the result is a smaller graph representation. In the latter case, by using weights, a single graph has no more influence than any other within the batch just due to its size (number of vertices or edges). This should be the goal, as larger graphs do not carry a higher contribution to the final error and thus, should be treated as having the same influence as any other graph. While this effect might appear subtle, it is important to note that without the use of batch normalization in CGCNN, the resulting formation energy MAE more than doubles. As such, changes to its calculation can bear significant changes in performance. 

When the collapse tolerance is increased, more vertices (points) are grouped together. When the number of samples is small (see Table \ref{tab:jarvis_results} vs. Table \ref{tab:jarvis_results2} for $e_{ij}, d_{ij}$ and exfoliation energy), despite the information loss that can occur through the averaging of edge features, the MAE produced is actually lower. We believe this is caused by regularization, resulting in better generalization on the test set. This is further backed by the difference in MAE decreasing as the size of the graph increases for properties with larger sample sizes (i.e. Jarvis-DFT formation energy).

Notably, both of these are not bound to a particular model. Instead, they address an issue with batch normalization applied to GNNs in general and introduce regularization in a new way that may be helpful, depending on the data being modeled.

While the distribution graph certainly provides benefits in its use, the formation of distribution graphs is currently contingent on vertices having the same outgoing degree. Graphs containing vertices with highly variable degrees will not benefit as significantly (or possibly not at all) when turned into a distribution graph. A possible approach for rectifying this is to allow for vertices with different, but very similar topologies, to be collapsed. The collapse tolerance addresses this case when the number of edges of two vertices is the same but the weights of the edges themselves vary slightly. Another mechanism that systematically allows for collapsing two vertices with different sets of edges is more difficult to design and would inevitably introduce some level of error through information loss that would likely be more significant than the collapse tolerance. A method for rectifying this would be applicable to graph neural networks regardless of their application.

The experiments contained in Subsection \ref{sec:col_tol_testing} and \ref{sec:k_sens_testing} to elucidate the effect of the two hyper-parameters of the distribution graph were in line with what would be expected based on the theoretical results of the PDD and measurement errors in crystal structures. For the line graphs in ALIGNN, there is a less justifiable reason for selecting $k$ and for using the collapse tolerance selected. First, it is unclear whether the value of $k$ determined using our method is too small or even too large given the additional angular information. Second, the collapse tolerance can be selected based on the size of measurement error in Angstroms or on mitigating floating point precision error. When the collapse tolerance is too large, the collapse of two vertices propagates to any derived line graph. This is fine if the topology in the line graph created by the vertices is the same, but should this not be the case the information loss produced from selecting either instead of keeping both would be significant. Notably, this is not a problem when the sample size of the training/testing data is small as seen in Table \ref{tab:jarvis_results}. Such a situation can be avoided by selecting an appropriate collapse tolerance or by using condition \textsc{(iv}.b\textsc{)} from Section \ref{sec:line_graphs}. For these reasons, however, these hyper-parameters still need to be assessed based on trial and error.

\section{Conclusion}
The DDG is a graph representation for periodic crystals that is invariant to rigid motion and unit cell choice. This representation reduces the size of the graph representation significantly compared to the commonly used crystal graph while improving prediction performance. By having fewer vertices and edges in the representation, we can speed up training and prediction time, as well as reduce memory requirements. Such an advantage is in line with the original intent of introducing machine learning methods into material property prediction.

We develop a graph representation of crystals based on the Pointwise Distance Distribution, a continuous and generically complete invariant of all periodic structures. While the graph cannot maintain continuity, it is sufficient for the finitely sized set of crystals within the T2 and Materials Project. Regardless, we find that the use of the PDD as the basis for our graph representation is a step towards the eventual goal of adapting the \textit{Crystal Isometry Space} for material property prediction. 

We adopt the two values from the PDD as hyper-parameters in the DDG, the collapse tolerance and $k$. The collapse tolerance can be tuned to increase or decrease the size of the graph in line with one's requirements. Further, it allows the DDG to be stable to atomic perturbations or small errors in measurement that can occur at the atomic level. Finally, we used properties of the PDD to help guide the selection of the hyper-parameter $k$, as well as explored how this affects the performance of prediction accuracy.

This DDG is generalized to the Distribution Graph, which can accommodate derived line graphs to include angles and dihedrals. We show this graph representation to outperform the crystal graph in accuracy and size on the Jarvis-DFT dataset across six material properties.

Distribution graphs can be used as a representation for any model that currently uses the Crystal Graph to reduce compute requirements and improve accuracy. Further, the modifications we described in the methodology for incorporating weighted vertices can be applied in general to GNNs to use vertex-weighted graphs.

\bibliographystyle{unsrt}  
\bibliography{references}  

\end{document}

%% file: math_commands.tex
\usepackage{amsmath,amsfonts,bm}

\def\eqref#1{equation~\ref{#1}}

\def\1{\bm{1}}

\DeclareMathAlphabet{\mathsfit}{\encodingdefault}{\sfdefault}{m}{sl}
\SetMathAlphabet{\mathsfit}{bold}{\encodingdefault}{\sfdefault}{bx}{n}

\newcommand{\R}{\mathbb{R}}

